 \documentclass[%preprintnumbers,%preprint,prd,%a4work,
 superscriptaddress,
twocolumn,showpacs, amsmath, amssymb, floatfix,
nofootinbib,wrapfloat byrevtex]{revtex4}

 \usepackage{amsmath,mathrsfs,graphicx,epstopdf,placeins,float}
 \usepackage{diagbox}
 \usepackage{pstricks}
 
 \newcommand{\beq}[1]{\begin{equation}\label{#1}}
 \newcommand{\eeq}{\end{equation}}
 \newcommand{\bea}[1]{\begin{eqnarray}\label{#1}}
 \newcommand{\eea}{\end{eqnarray}}

\begin{document}

 \title{$d_{\ell}(z)$ and BAO in the emergent gravity and the dark universe}
 \author{Ding-fang Zeng}
 \email{dfzeng@bjut.edu.cn}
 \affiliation{Theoretical Physics Division, College of Applied Sciences, Beijing University of Technology}   
 \begin{abstract}
 We illustrate that $\Lambda$MOND cosmology following from E. Verlinde's emergent gravity idea which contains only constant dark energy and baryonic matters governed by linear inverse gravitation forces at and beyond galaxy scales fit with the luminosity distance v.s. redshift relationship, i.e. $d_\ell(z)$ of type Ia supernovae equally well as the standard $\Lambda$CDM cosmology does. But in a rather broad and reasonable parameter space, $\Lambda$MOND gives too strong baryon acoustic oscillation, i.e. BAO signals on the matter power spectrum contradicting with observations from various galaxy survey and counting experiments.
 \end{abstract}
 \pacs{04.20.Cv, 98.65Dx, 95.35+d, 95.36+x}
 \maketitle

In a last month's work \cite{Verlinde2016}, basing on insights from string theory, black hole physics and quantum information theory, Eric. Verlinde argues that the dark gravity effects observed in galaxies and clusters conventionally attributed to dark matters could be accounted for by the modified newtonian dynamics (MOND hereafter) following from the emergence feature of gravitation and space-time itself. Although Verlinde does not quotient concrete method for modifying the newton gravitation theory, so there is big arbitrarinesses in its prediction possibilities. His idea attracts much attention \cite{vLciteKB,vLciteLP,vLciteBetal,vLciteIorio,vLciteBueno,vLciteDGN,vLciteEttori,vLciteBM} as well as criticisms \cite{MONDcriticism} due to its potential of kicking the longly non-measured dark matter contents out of our knowledge menus using first principle of quantum gravitation theories. 

Historical research works \cite{Milgrom1983,MOND2,MOND3,MOND4,MOND5} indicate that if at and beyond galaxy scales the square-inverse feature of newton gravitation is enhanced appropriately, e.g. the most simple way of enhancing to linear-inverse laws, then the galaxy rotation curves could be flattened properly just as observation requires. Verlinde argues that this square-inverse to linear-inverse transition should occur for general gravitation systems as long as their characteristic acceleration goes below the magic value,
\beq{}
a^\mathrm{acc}_\mathrm{magic}=\frac{1}{6}H_0c
\eeq
where $H_0$ is today's hubble parameter. Since in a matter dominated universe $a\propto t^\frac{2}{3}$, $\ddot{a}\propto-a^{-2}$, we expect that the universe as a gravitation system should also experience this square-inverse to linear inverse transition as its scale factor grows beyond
\beq{}
a^\mathrm{s.f.}_\mathrm{magic}\approx\big(\frac{a_0^2}{6}\big)^\frac{1}{2}
\eeq   
This reasoning immediately brings us two questions urgently. The first is, if the universe consists only of normal matters controlled by this modified gravitation theory and constant dark energy($\Lambda$MOND here after), then could its late time expansion features still be similar to that observed in the type Ia supernovae's distance-redshift relationship? The second is, could this $\Lambda$MOND model reproduce the $\Lambda$ plus Cold Dark Matter($\Lambda$CDM here after) model's beautiful prediction of large scale structure's evolution and growth or not? At first glance, we may think that this two questions may not be answered properly before concrete MOND formulation thus well-established new gravitational field equation following from the emergent idea is spelled out. However, we will show in the following that this is not the case.

\begin{figure}[h]
\rule{8mm}{0pt}\includegraphics[clip=true,bb=0 30 235 221]{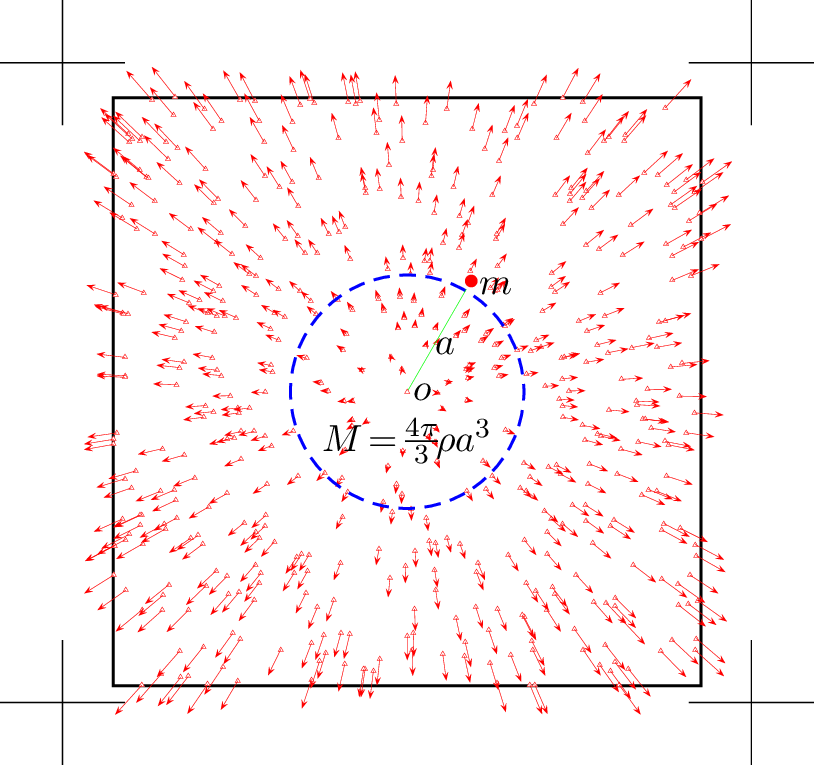}
\caption{$o$ is arbitrary observer, while $m$ an arbitrary co-moving test particle in an isotropic and homogeneous expanding universe. The energy conserving condition obeyed by $m$ and the mass/energy contents inside the spherical region $M$ centered on $o$ simply has the form $\frac{1}{2}m\dot{a}^2r^2-\frac{GMm}{ar}=\tilde{k}$, with $\tilde{k}$ a constant determined by initial conditions.}
\label{figEconservation}
\end{figure}
According to the standard textbook of S. Weinberg, in an isotropic and homogeneous universe
\beq{}
ds^2=-dt^2+a^2(t)\big[\frac{dr^2}{1-kr^2}+r^2d\theta^2+r^2\sin^2\!\theta\,d\phi^2\big]
\eeq
simple newton mechanic and energy conservation laws are sufficient in determining dynamics of the scale factor $a(t)$. Referring FIG.\ref{figEconservation} and captions there, according to which the conventional Friedman equation could be derived as follows
\beq{}
\frac{1}{2}m\dot{a}^2r^2-\frac{4\pi G}{3}\frac{(\rho_m+\rho_\Lambda) a^3r^3}{ar}=\tilde{k}
\label{eqNewtonmechanic}
\eeq
\beq{}
\Rightarrow\frac{\dot{a}^2}{a^2}-\frac{k}{a^2}=\frac{8\pi G}{3}(\rho_m+\rho_\Lambda)
\label{eqFriedmann}
\eeq
while the energy conservation requires
\beq{}
\rho_ma^3=\mathrm{const.1},~\rho_\Lambda=\mathrm{const.2}
\label{eqEconservation}
\eeq
Equations \eqref{eqFriedmann} and  \eqref{eqEconservation} constitute the full set of dynamic constraints for $a(t)$.

Now, let us apply the above method to the MOND theory of E. Verlinde in which the matter-matter attractive force has linearly inverse law at scales beyond galaxies but the matter-dark energy force still satisfies the square inverse law, 
\beq{}
V_\mathrm{mm}=-\frac{1}{\epsilon}\frac{GM_\mathrm{m}m}{(ar)^\epsilon}\Leftarrow F=-\nabla V=-\frac{GM_\mathrm{m}m}{(ar)^{1+\epsilon}}
\eeq
\beq{}
V_{\mathrm{m}\Lambda}=-\frac{GM_{\mathrm{m}\Lambda}m}{ar}\Leftarrow F=-\nabla V=-\frac{GM_{\mathrm{m}\Lambda}m}{a^2r^2}
\eeq
\beq{}
\epsilon=\frac{1}{2}+\frac{1}{2}\tanh\big(\frac{a_0}{a}-\frac{a_0}{a^\mathrm{s.f.}_\mathrm{magic}-a}\big)\theta(a^\mathrm{s.f.}_\mathrm{magic}-a)
\label{epsfunc}
\eeq
We introduce a simple $\epsilon(a)$ function here to implement the goal of changing the early time square-inverse law to the later time linear inverse law smoothly, where $\theta$ is the usual heaviside step function featured by $\theta(x\leqslant0)=0$, $\theta(0<x)=1$.

Substituting the above two potential formulas into the conservation equation \eqref{eqNewtonmechanic}, what we get will become
\beq{}
\frac{\dot{a}^2}{a^2}-\frac{k}{a^2}=\frac{8\pi G}{3}\big[\rho_m \frac{ar/\epsilon}{(ar)^\epsilon}+\rho_\Lambda\big]
\label{eqMNDFriedman}
\eeq
In the case of $k=0$ and the matter dominated era, the function $a(t)$ could be explicitly solved out
\beq{}
a_\mathrm{mond}(t)=\big(\frac{8\pi G}{3}\frac{r/\epsilon}{r^\epsilon}\big)^\frac{1}{2+\epsilon}\frac{2+\epsilon}{2}t^\frac{2}{2+\epsilon}|_{\epsilon\rightarrow0}
\label{amondSol}
\eeq
Considering the fact that $\dot{a}^{\epsilon\rightarrow0}_\mathrm{mond}$ $\propto t^0$ $\rightarrow\mathrm{const.}$, while in the standard Einstein/Newton cosmologies $\dot{a}\propto t^{-\frac{1}{3}}\rightarrow0$, it is very surprising that the strengthened gravitation force does not lead to strengthened deceleration and re-collapsing evolutions of the universe. Of course, when the constant dark energy is included, there existing re-collapses or not will be relevant with the relative weight of $\rho_{m0}$, $\rho_{\Lambda0}$ and $k$. Any way, this fact suggests us that this simple $\Lambda$MOND model may have radically different late time expansion features relative to the standard $\Lambda$CDM model.

\begin{figure}[h]
\includegraphics[scale=0.47]{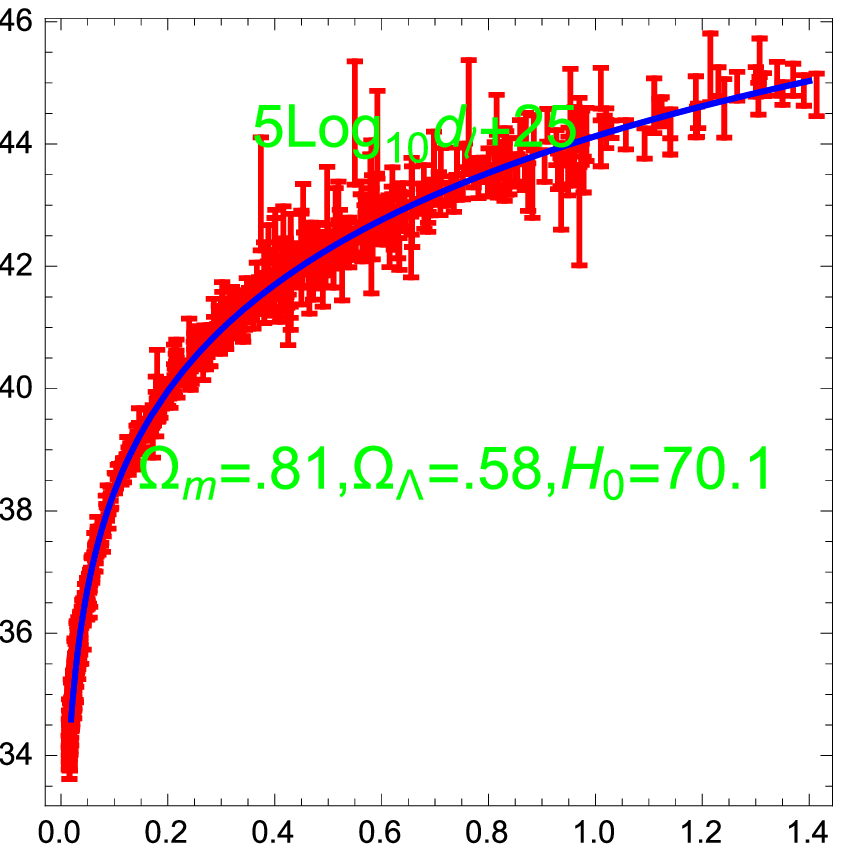}
\includegraphics[scale=0.47]{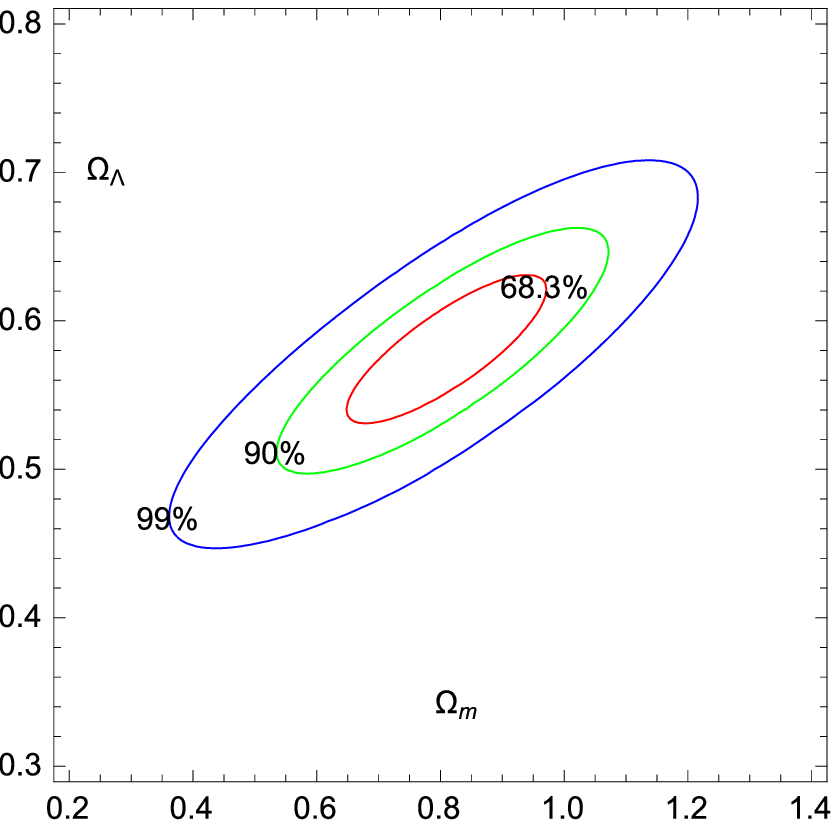}\\
\includegraphics[scale=0.47]{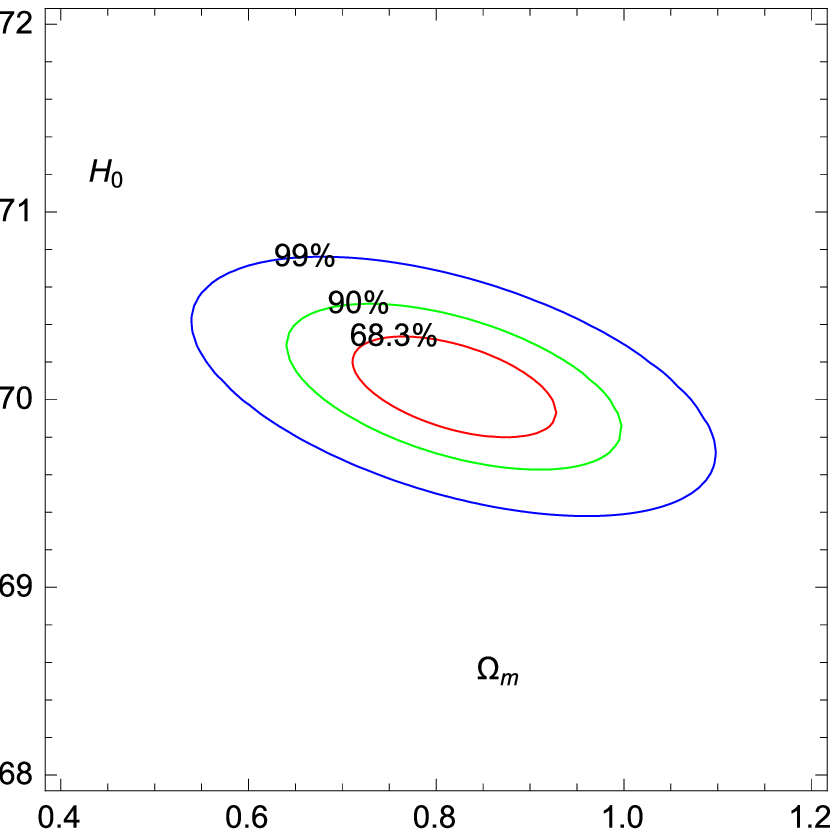}
\includegraphics[scale=0.47]{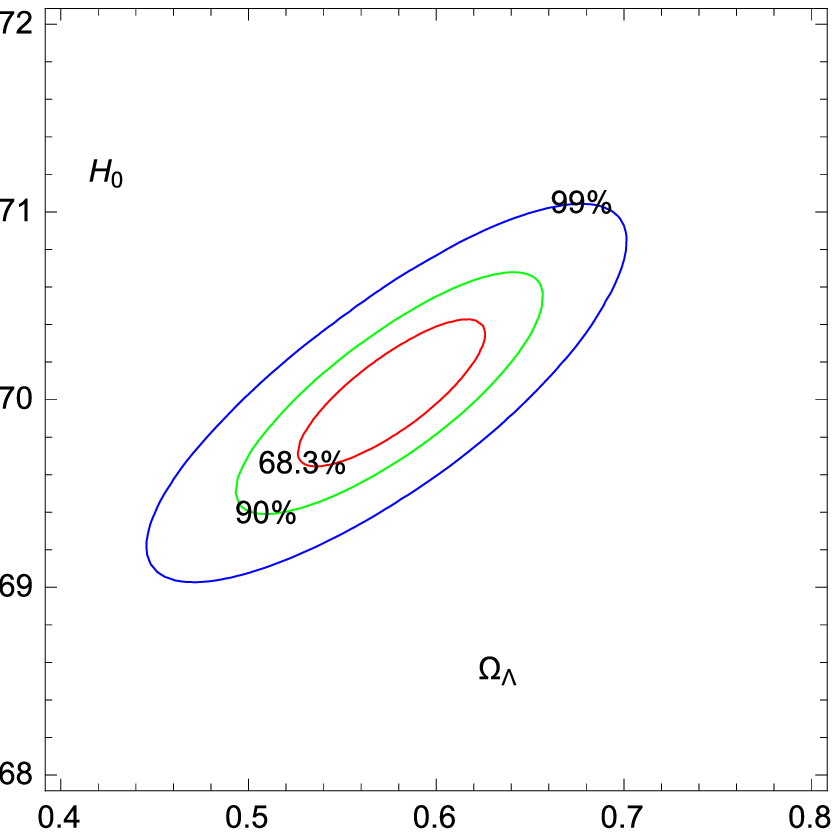}
\caption{Best fitting the 580 data points of the SCP Union2.1 \cite{SCP2011} with $\{\Omega_m,\Omega_\Lambda,H_0\}$ as free free parameters in the $\Lambda$MOND cosmology by the simple $\chi^2$-minimization method.}
\label{figEconservation}
\end{figure}

Using equation \eqref{eqMNDFriedman} and the standard definition in conventional supernovae data analysis, we can derive out the luminosity distance v.s. redshift relation in the $\Lambda$MOND model,
\beq{}
d_\ell(z)=\frac{a_0^2}{a}\frac{c/(a_0H_0)}{|\Omega_k|^\frac{1}{2}}\mathrm{sinn}\big[|\Omega_k|^\frac{1}{2}\int_0^z\frac{d\zeta}{H(\zeta)/H_0}\big]
\eeq
\beq{}
H^2(\zeta)=H_0^2\big[\frac{\Omega_ma_0^3}{a^3}\big(\frac{a}{a_0}\big)^{1-\epsilon}+\Omega_\Lambda-\frac{\Omega_ka_0^2}{a^2}\big]
\eeq
where we defined $\frac{a_0}{a}\equiv\zeta+1$,~$\mathrm{sinn}[x]$ $=$ $\sin[x]$, $x$, $\sinh[x]$ as $\Omega_k>0$,$=0$ and $<0$ respectively. We are very lucky that the annoying factors $\frac{r/\epsilon}{r^\epsilon}$ appearing in \eqref{eqMNDFriedman} could be simply absorbed into the definition of $\Omega_m$, so that we can now safely set $\epsilon=0$ here. Now using observational data complied in the SCP Union2.1 \cite{SCP2011} and minimizations of the following $\chi^2$-function
\bea{}
\chi^2=\sum_i\frac{[m_\mathrm{th}(z_i,\Omega_m,H_0,\cdots)-m_\mathrm{ex}(z_i)]^2}{\sigma_i^2}
\\
m_\mathrm{th}(z,\cdots)\equiv5\log_{10}[d_\ell(z,\cdots)/\mathrm{mpc}]+25
\nonumber
\eea
we find that the three parameter $\{\Omega_m,\Omega_\Lambda,H_0\}$ $\Lambda$MOND model fit with the observational data equally well with the standard $\Lambda$CDM model. But with radically different best fitting parameters
\bea{}
&&\hspace{-5mm}\Lambda\mathrm{MOND}:~\Omega_m=0.81,~\Omega_\Lambda=0.58
,H_0=\frac{70.1\mathrm{km}}{\mathrm{s}\!\cdot\!\mathrm{mpc}}
\\
&&\hspace{-3mm}\Lambda\mathrm{CDM}:~\Omega_m=0.29,~\Omega_\Lambda=0.76
,H_0=\frac{70.2\mathrm{km}}{\mathrm{s}\!\cdot\!\mathrm{mpc}}
\eea
The former has $\chi^2=562.313$, while the latter has $\chi^2=562.40$. Just as we pointed out under equation \eqref{amondSol} that the strengthened gravitational force in this model between matters does not make the acceleration of the universe difficult. Instead they make such accelerations more easier so that more less dark energy is need in accomplishing the observed acceleration!

The most big difficult a cosmological model without non-baryonic dark matter may encounter is that, it may lead to too strong baryonic acoustic oscillation (BAO) signal on the power spectrum of matter distributions such as those observed typically in 2dFGRS \cite{2dFGRS} and SDSS \cite{BAO0501,BAO0608} galaxy survey and counting experiments. Recalling that in the standard cosmological perturbation theory \cite{DodelsonTextbook}
\beq{}
ds^2=a^2(\eta)\big[-(1+2\Psi)d\eta^2+(1+2\Phi)\delta_{ij}dx^idx^j\big]
\eeq
\bea{}
&&\hspace{-5mm}k^2\Phi+3\frac{\dot{a}}{a}\big(\dot{\phi}-\Psi\frac{\dot{a}}{a}\big)
=4\pi Ga^2\big(\rho_{\!_D}\delta_{\!_D}+
\label{Ein00pert}\\
&&\hspace{15mm}\rho_{\!_B}\delta_{\!_B}+4\rho_\gamma\Theta_0+4\rho_\nu\mathcal{N}_0\big)
\nonumber
\eea
\beq{}
k^2(\Phi+\Psi)=-32\pi Ga^2\big(\rho_\gamma\Theta_2+\rho_\nu\mathcal{N}_2\big)
\label{Einijpert}
\eeq
The last two equations follow from Fourier transformations of the linearised Einstein equation. Quantities on their right hand side are just the 1st and 2nd multipole expansions of the corresponding particle's statistical distribution
\bea{}
\mathrm{photon}:&&\hspace{-3mm}f_\Theta=\big[e^{\frac{p}{T[1+\Theta(\vec{x},\hat{p},t)]}}-1\big]^{-1}
\\
\mathrm{nutrino}:&&\hspace{-3mm}f_\mathcal{N}=\big[e^{\frac{p}{T[1+\mathcal{N}(\vec{x},\hat{p},t)]}}+1\big]^{-1}
\\
\mathrm{dark~matter}:&&\hspace{-3mm}f_{\delta{D}}=\cdots,
\\
\mathrm{baryon~matter}:&&\hspace{-3mm}f_{\delta{B}}=\cdots
\eea
\bea{}
&&\hspace{-3mm}\Theta_\ell(\vec{x},t)\equiv-\frac{i^\ell}{2}\!\int_{-1}^1\!d\cos\!\theta\,\Theta(\vec{x},\cos\theta,t)P_\ell(\cos\theta)
\\
&&\hspace{-3mm}\mathcal{N}_0,~\mathcal{N}_2,~\delta_{\scriptscriptstyle D}\equiv\delta_{\scriptscriptstyle D0},~\delta_{\scriptscriptstyle B}\equiv\delta_{\scriptscriptstyle D0}~\mathrm{similarly~defined}
\nonumber
\eea
Unlike $\Psi$ and $\Phi$, all these multipole's evolution is controlled directly by the Boltzmann instead of Einstein equation
\beq{}
\frac{df}{dt}=C[f(\vec{p})]
\eeq
The concrete form $C$ depends on the particle type and their mutual interactions. At first two levels, the component equation relevant with the baryon acoustic oscillation reads
\beq{}
\dot{\Theta}_0+k\Theta_1=-\dot{\Phi},~\tau(\eta)\equiv-n_e\sigma_Ta
\eeq
\beq{}
\dot{\Theta}_1-\frac{k}{3}\Theta_0=\frac{k}{3}\Psi+\dot{\tau}\big[\Theta_1-\frac{iv_{\scriptscriptstyle B}}{3}\big]
\eeq
\beq{}
v_b=-3i\Theta_1+\frac{R}{\dot{\tau}}\big(\dot{v}_{\scriptscriptstyle B}+\frac{\dot{a}}{a}v_{\scriptscriptstyle B}+ik\Psi\big)
\eeq
Under the so called tightly-coupling limit, Hu and Sugiyama show \cite{WayneHuSugiyama1995} that this equation array has simple oscillation solution
\bea{}
\Theta_0(\eta)+\Phi(\eta)=[\Theta_0(\eta)+\Phi(\eta)]\cos(kr_s)
\label{tightCouplingLimit}\\
+\frac{k}{\sqrt{3}}\int_0^\eta\!dy\big[\Phi(y)-\Psi(y)\big]\sin\{k[r_s(\eta)-r_s(y)]\}
\nonumber
\eea
\beq{}
r_s\equiv\int_0^\eta dyc_s(y),~c_s\equiv(3+3R)^{-\frac{1}{2}},~R\equiv\frac{3\rho_{\scriptscriptstyle B}}{4\rho_\gamma}
\eeq
This is just the baryon acoustic oscillation. It originates from the sound mode oscillation of the relativistic plasma in the early universe. At redshift $z\approx1000$, the recombination occurs so that the big bang plasma becomes a neutral gas and the oscillation stops propagating any further. But periodic spatial inhomogeneity feature it brings continues to exist and evolves to the present time. In the standard CDM model, the baryonic and non-baryonic dark matters coexist even before the recombination, with the former to latter ratio equates about $\frac{1}{5}$. Since dark matters do not participate in the sound wave oscillation, the strength of baryonic acoustic oscillation signals on the power spectrum of total matters observed in the late time universe is very small. In cosmological models such as $\Lambda$MOND where non-baryonic dark matters do not exist at all, to explain the smallness of of this signal is the main challenge. 

\begin{figure}[h]
\includegraphics[scale=0.5]{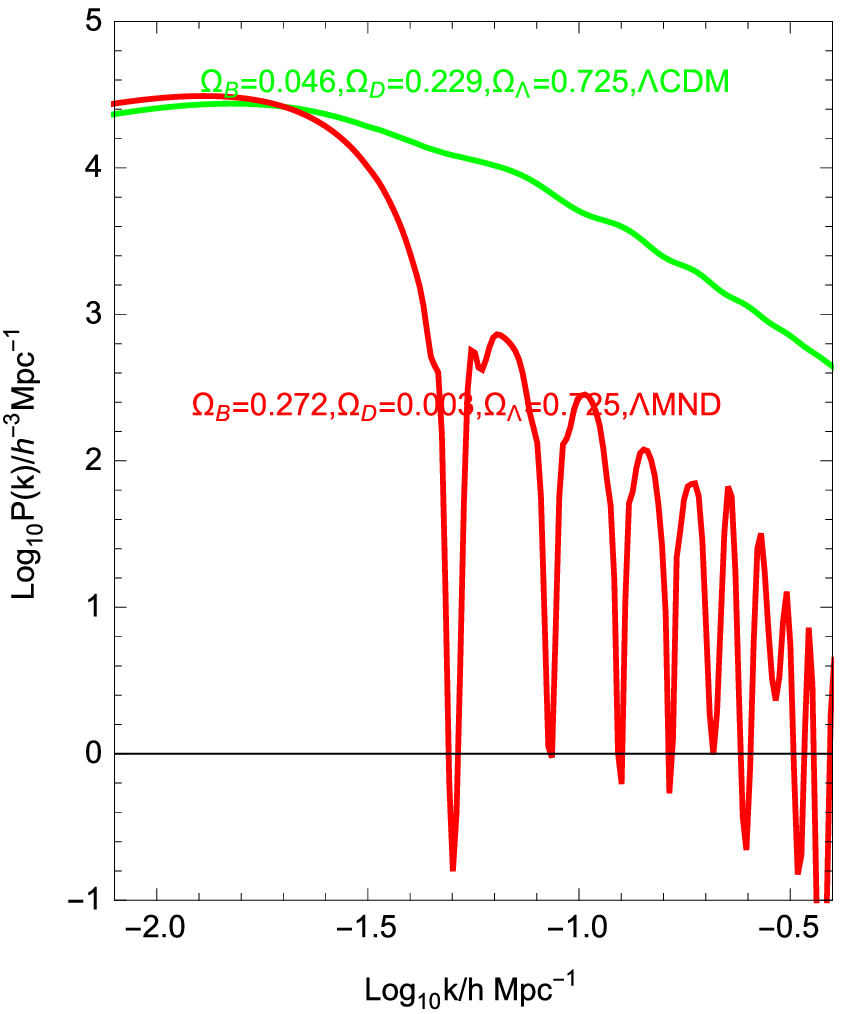}
\includegraphics[scale=0.5,clip=true,bb=20 0 240 291]{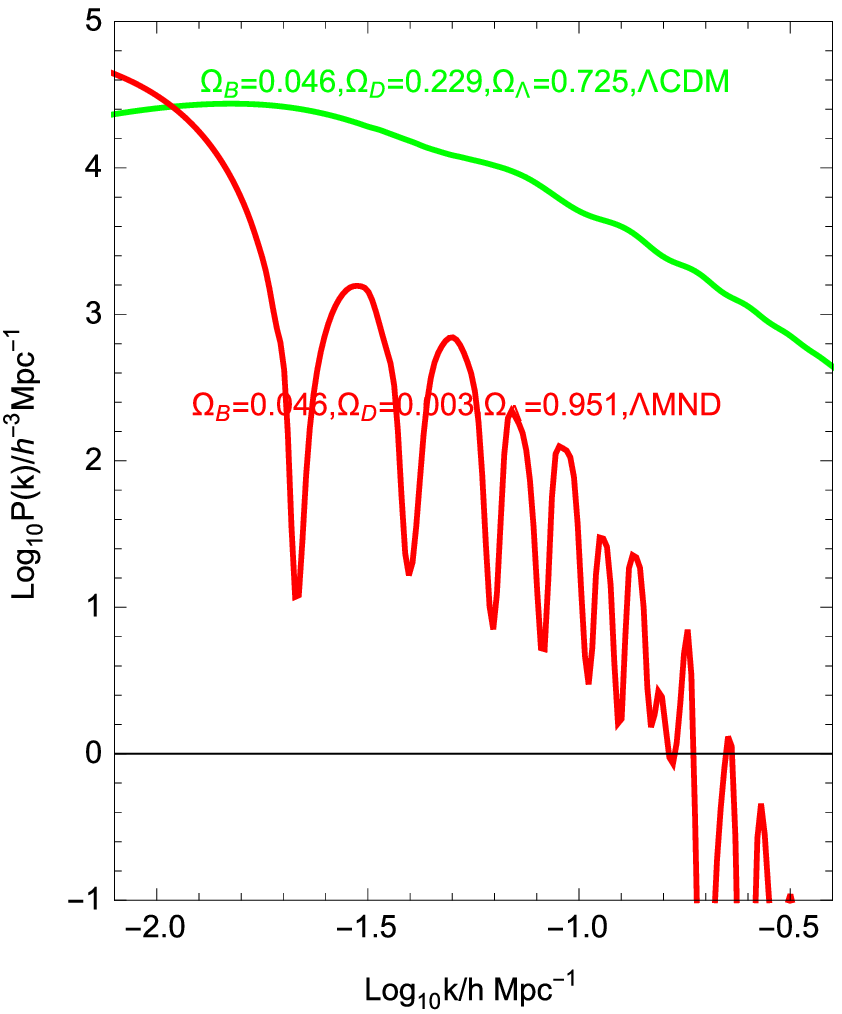}
\caption{In standard $\Lambda$CDM model (green line), the BAO signal manifests only as small wiggles on the matter power spectrum. But in the $\Lambda$MOND model (red line), this signal manifests as strong oscillations of the power spectrum line. In the left panel, we replace the dark matter of $\Lambda$CDM with baryonic matters in $\Lambda$MOND, while in the right panel, we replace it with dark energy. In both panels, $H_0$ is set as $70\mathrm{km}/(\mathrm{s}\cdot\mathrm{Mpc})$.}
\label{figBAOsignalA}
\end{figure}
We explore in the following if modifying the standard Friedmann equation into the form \eqref{epsfunc}-\eqref{eqMNDFriedman} and letting all linear perturbations evolve in this background the same way as the usual Einstein-Botzmann formulae require brings us suppressions of the BAO signal as required by observations. Our logic is, although we do not know what the full gravitational field equation grows like in the framework of emergent idea, in an exactly isotropic and homogeneous universe its key features are captured by \eqref{epsfunc}-\eqref{eqMNDFriedman}, while its linear perturbation, as long as being second order partial differential equations, would then not deviate from the Einstein perturbations too much. 

\begin{figure}[h]
\includegraphics[scale=0.5]{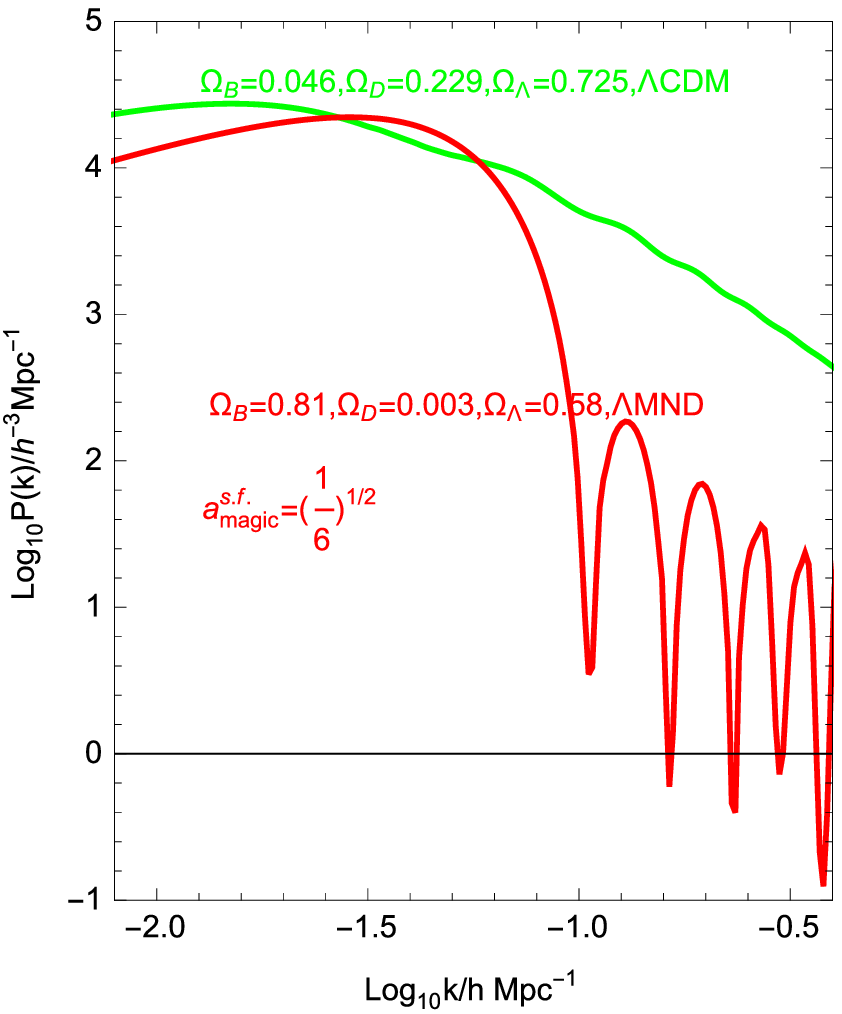}
\includegraphics[scale=0.5,clip=true,bb=20 0 240 291]{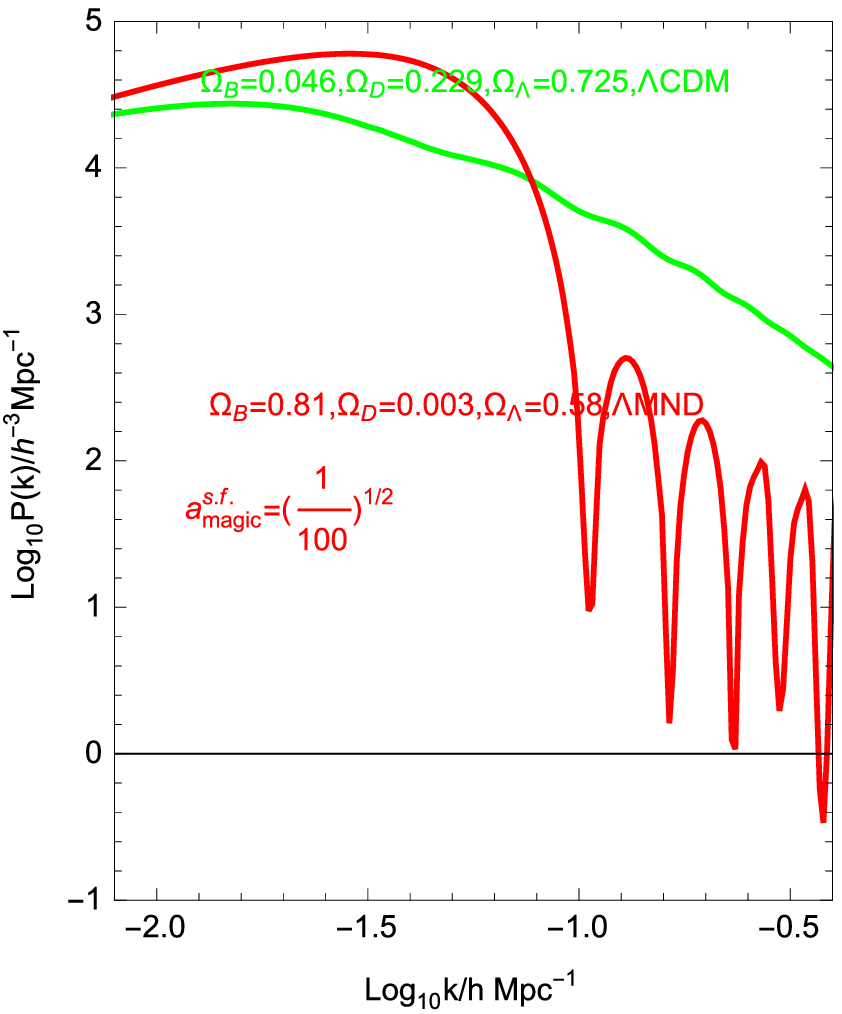}
\caption{Similar as FIG.\ref{figBAOsignalA}, but we used $\Omega_m,\Omega_\Lambda$ parameters following from best fittings of the $d_\ell(z)$ relation of type Ia supernovae. In the left panel we let $a^\mathrm{s.f.}_\mathrm{magic}=(1/6)^{1/2}$ as required by E. Verlinde's argument. While in the right, we tries to set it to more smaller values.}
\label{figBAOsignalB}
\end{figure}

Under this logic, we integrate the whole system of Einstein-Boltzmann differential equations by the standard code of CAMB \cite{CAMBcode2000}. The results is displayed in FIG.\ref{figBAOsignalA} and \ref{figBAOsignalB} explicitly. From FIG.\ref{figBAOsignalA}, we easily see that, replacing the conventional cold dark matter with either baryonic matter of MOND or dark energy of cosmological constant both bring us too strong BAO signals on the matter power spectrums measured observationally. Nevertheless, in the former case, $\Lambda$MOND and $\Lambda$CDM has approximately the same first peak positions on the power spectrum. This is expectable because it could be proved analytically basing on the tight-coupling approximation \eqref{tightCouplingLimit} whose validity has no relevance with assumptions of the $\Lambda$MOND model.

To obtain a suppressed BAO signal, we try in FIG.\ref{figBAOsignalB} using $\Omega_m,\Omega_\Lambda$ parameters following from best fittings of the $d_\ell(z)$ relation of type Ia supernovae in the previous section, where spatial curvatures contribute remarkably heavier to the energy contents of the universe. However, even when we let the magic scale factor be a tuneable parameter, we do not obtain the required results. This means that, new mechanisms must be find to suppress the BAO signal in this $\Lambda$MOND cosmology to make it a competing model of $\Lambda$CDM.

Conclusion: we derive out dynamic equations controlling the evolution of scale factors in a simple $\Lambda$MOND cosmology which contains only constant dark energy and baryonic matters governed by linear inverse gravitation forces at and beyond galaxy scales. We find that the model fit with observational data type Ia supernovae's luminosity distance v.s. redshift relationship equally well with the standard $\Lambda$CDM model does. However, since no dark matter is assumed, the model predicts too strong baryonic acoustic oscillation signals on the matter power spectrum than the standard $\Lambda$CDM does. Nevertheless, $\Lambda$MOND has the same position of first BAO peak as $\Lambda$CDM does if we replace dark matters in the latter with baryonic matters in the former. So a reasonable mechanism to suppress the strength of BAO signals may be the most urgent ingredient of $\Lambda$MOND in its road of growing into competing models of $\Lambda$CDM.

\section*{Acknowledgements}
This work is supported by Beijing Municipal Natural Science Foundation, Grant No. Z2006015201001 and partly by the Open Project Program of State Key Laboratory of Theoretical Physics, Institute of Theoretical Physics, Chinese Academy of Sciences, China.

\end{document}